\documentclass[namedreferences,hyperref,optionalrh]{spr-sola}
\usepackage{graphicx}        
\usepackage{color}           
\usepackage{amsmath} 

\chardef\us=`\_

\begin{document}

\begin{frontmatter}
\title{A method of Extracting Flat Field from Real Time Solar Observation Data}

\author[addressref={aff1,aff2}]{\inits{Y.}\fnm{YaHui}~\snm{Jin}\orcid{0000-0001-9917-2926}}


\author[addressref=aff1,corref,email={liuhui@ynao.ac.cn}]{\inits{H.}\fnm{Hui}~\snm{Liu}\orcid{987-654-3210}}


\author[addressref=aff1]{\inits{K.}\fnm{KaiFan}~\snm{Ji}\orcid{987-654-3210}}

\author[addressref=aff1]
{\inits{Z.}\fnm{ZhenYu}~\snm{Jin}\orcid{987-654-3210}}

\author[addressref={aff1,aff2}]
{\inits{W.}\fnm{WeiJie}~\snm{Meng}\orcid{987-654-3210}}

\address[id=aff1]{Yunnan Observatories, Chinese Academy of Sciences, Kunming 650011, China}
\address[id=aff2]{University of Chinese Academy of Sciences, Beijing 100049, China}

\runningauthor{Author-a et al.}
\runningtitle{\textit{Solar Physics} Example Article}

\begin{abstract}
Existing methods for obtaining flat field rely on observed data collected under specific observation conditions to determine the flat field. However, the telescope pointing and the column-fixed pattern noise of the CMOS detector change during actual observations, causing residual signals in real-time observation data after flat field correction, such as interference fringes and column-fixed pattern noise. In actual observations, the slight wobble of the telescope caused by the wind leads to shifts in the observed data. In this paper, a method of extracting the flat field from the real-time solar observation data is proposed. Firstly, the average flat field obtained by multi-frame averaging is used as the initial value. A set of real-time observation data is input into the KLL method to calculate the correction amount for the average flat field. Secondly, the average flat field is corrected using the calculated correction amount to obtain the real flat field for the current observation conditions. To overcome the residual solar structures caused by atmospheric turbulence in the correction amount, real-time observation data are grouped to calculate the correction amounts. These residual solar structures are suppressed by averaging multiple groups, improving the accuracy of the correction amount. The test results from space and ground-based simulated data demonstrate that our method can effectively calculate the correction amount for the average flat field. The NVST 10830\:\AA/H$\alpha$ data were also tested. High-resolution reconstruction confirms that the correction amount effectively corrects the average flat field to obtain the real flat field for the current observation conditions. Our method works for chromosphere and photosphere data.
 
\end{abstract}
\end{frontmatter}

\section{Introduction}
     \label{S-Introduction}
Flat field correction is an important part of astronomical observation data processing, especially in high-resolution ground-based solar observations, where the accuracy of the flat field directly affects the reconstruction results of observed data. Therefore, all large-aperture ground-based solar telescopes, such as the Swedish Solar Telescope (SST), Goode Solar Telescope (GST), New Vacuum Solar Telescope (NVST), and GREGO Solar Telescope \citep{yang2022gst}, have made improving the flat field quality an important task. 

\begin{figure}    
\label{image1}
\centerline{\includegraphics[width=0.9\textwidth]{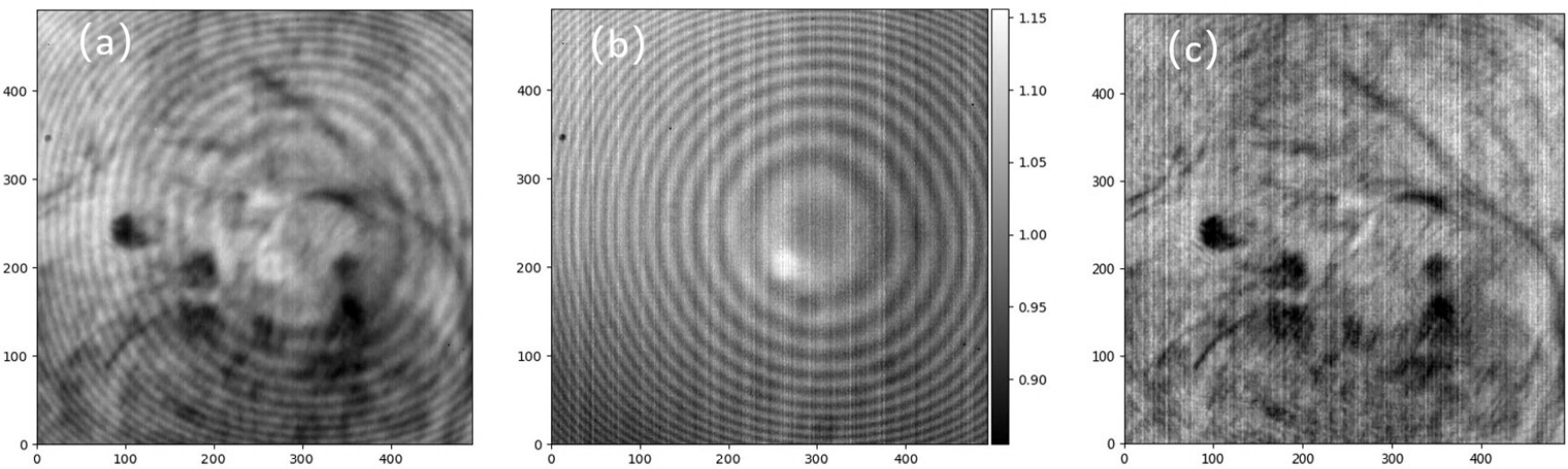}}
\caption{Panel (a) is NVST/He I 10830\:\AA\ observed data (the observed time is 13:42:37 UT on June 5, 2023), panel (b) is the average flat field (the observed time is 07:35:26 UT on June 5, 2023), and panel (c) is the high-resolution data after correction of the average flat field. It can be easily seen that the high-resolution data still contain residual signals after the average flat field correction, such as interference fringes, column-fixed pattern noise. The reconstruction method is non-rigid alignment \citep{liu2022high}}.
\end{figure}

Flat field obtaining methods can be divided into two categories: one is the multi-frame averaging method for small field of view (FOV) observations, which moves the telescope near the solar quiet region and then averages multiple images to obtain the flat field (called average flat field). This method is simple and has high accuracy. The root mean square (RMS) of pixel-to-pixel intensity variation in the flat field is about 0.1\% \citep{moran1992photometric,yang2003high}. The other is the KLL method \citep{kuhn1991gain} for large FOV observations, which was improved by \citet{chae2004flat}. The KLL method records multiple images of solar disk with different parts of the detector by changing the telescope’s pointing. The obtained images are used to calculate the flat field by iteration with some specific algorithm \citep{li2021methodology}. At present, it has been widely used in space or ground-based full-disk observations \citep{denker1999synoptic,boerner2012initial,fang2013new}. These methods all rely on a flat field observation pattern to move the telescope and obtain specialized flat field observed data to calculate the flat field. However, the observation conditions change during actual observations. In particular, the change in the telescope’s pointing alters the relative angle between incident light and the detector's target surface, which in turn affects the intensity of interference fringes in the observed data. In addition, the column-fixed pattern noise of the CMOS detector changes over time \citep{wang20161,qiu2021}. Therefore, even if these methods can obtain a highly accurate flat field, it is still not the real flat field for the current observation conditions. When the flat field obtained by these methods is used for flat field correction of real-time observation data, the corrected observed data still contains residual signals, such as interference fringes and column-fixed pattern noise. These residual signals are further amplified in subsequent high-resolution reconstruction \citep{liu2021new}, as shown in Figure~\ref{image1}. Therefore, we need a method for obtaining flat field that can adapt to changes in observation conditions. 

In small FOV ground-based solar observations, the slight wobble of the telescope caused by the wind leads to shifts in the observed data. We find that displacement characteristics of the observed data meet the requirements for extracting a flat field using the KLL method. Moreover, these observed data are obtained from real-time observations. Therefore, we can input a set of real-time observation data into the KLL method to extract the real flat field for the current observation conditions. However, it is very difficult to extract the real flat field directly from real-time observation data. Although there is a residual signal in the real-time observation data after the average flat field correction, it is still a highly accurate flat field. Therefore, we regard real flat field of real-time observation data as the product of the average flat field and the correction amount. The average flat field is used as the initial value, and the correction amount for the average flat field is dynamically calculated from the real-time observation data using the KLL method. Finally, the calculated correction amount is used to correct the average flat field to obtain the real flat field for the current observation conditions. However, the calculated correction amount may have residual significant solar structures, which will reduce the accuracy of the correction amount. This is because small FOV ground-based observed data are seriously distorted by atmospheric turbulence, which violates the KLL method’s assumption that the solar intensity remains stable for a short time. Due to the random distortion of the observed data caused by atmospheric turbulence, the residual solar structures in the correction amount also exhibit random characteristics. Therefore, we can divide a large number of real-time observation data into multiple groups and use the KLL method to calculate the correction amount from each group. Finally, the correction amounts from multiple groups are averaged to suppress the residual solar structures, thereby improving the accuracy of the correction amount.         

In this paper, a method of extracting flat field from real-time solar observation data is proposed. Firstly, a large number of real-time observation data are divided into multiple groups. Secondly, the average flat field is used as the initial value, and the correction amount for the average flat field is calculated from each group of real-time observation data using the KLL method. Thirdly, multiple groups of correction amount are averaged to suppress residual solar structures, thereby improving the accuracy of the correction amount. Finally, the calculated correction amount is used to correct the average flat field to obtain the real flat field for the current observation conditions. The test results of space and ground-based simulated data demonstrate that our method can effectively calculate the correction amount for the average flat field. The NVST He I 10830\:\AA/H$\alpha$ data were also used for testing. High-resolution reconstruction confirms that the calculated correction amount effectively corrects the average flat field to obtain the real flat field for the current observation conditions.

The contents of this paper are organized as follows: In Section 2, the space and ground-based simulated data, NVST He I 10830 Å/H$\alpha$ observation data are described. In Section 3, the proposed method is introduced in detail. In Section 4, simulated data is used to test the effectiveness of our method for calculating the average flat field correction amount. In Section 5, NVST He I 10830 Å/H$\alpha$ observed data is used to test the proposed method. In Sections 6, and 7 are discussed and summarized.

\section{Data}
\subsection{Space/Ground-based Simulated Data} 
    \label{S-sim}
In order to test the validity of our method in calculating the correction amount for the average flat field, we constructed two sets of simulated data: (1) The 200 frames of space observed simulated data not affected by atmospheric turbulence. (2) The 100 sets of ground-based observed simulated data (200 frames each) affected by atmospheric turbulence.
   
\begin{figure}    
\centerline{\includegraphics[width=0.9\textwidth]{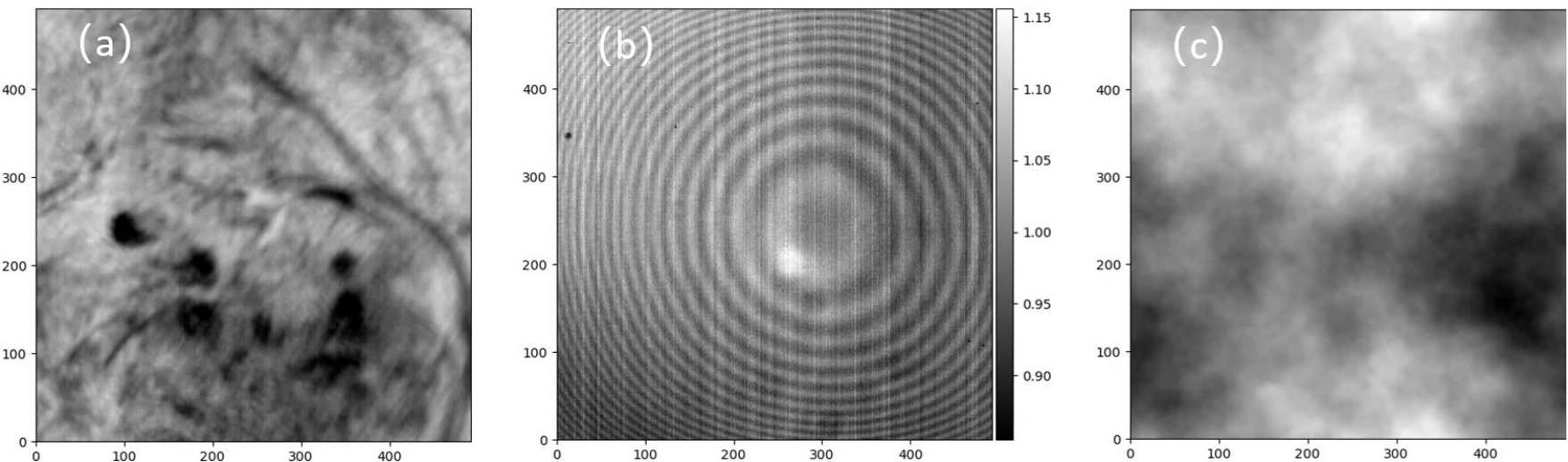}}
\caption{Panel (a) is the pre-processed NVST/He I 10830\:\AA\ observation data, which is used as the observation object (the observation time is 13:42:37 seconds on June 5, 2023), panel (b) is the average flat field (the observation time is 07:35:26 UT on June 5, 2023), which is used as the real flat field of the simulated data, and panel (c) is one of the randomly generated phase screens using the power spectrum inversion method \citep{mcglamery}.}
\label{moni}
\end{figure}

The simulated data (1) is constructed as follows: Firstly, a frame of flat dark field preprocessed NVST/He I 10830\:\AA\ observed data (the observed time is 13:42:37 UT on June 5, 2023) is used as the solar observation object (Figure~\ref{moni}a). The average flat field (the observation time is 07:35:26 UT on June 5, 2023) is used as the real flat field (Figure~\ref{moni}b) of the space observed simulated data. Secondly, 200 groups of relative displacements (following a Gaussian distribution, ranging from 5 sub-pixels) are randomly generated to simulate the slight wobble of the telescope caused by the wind during actual observation. Finally, the displacements are integrated into the solar observation object (Figure~\ref{moni}a), and the displaced observation object is multiplied by the real flat field (Figure~\ref{moni}b) to generate 200 frames of space observed simulated data. 

The simulated data (2) is constructed as follows: Firstly, the simulated data (2) and the simulated data (1) use the same solar observation object (Figure~\ref{moni}a) and the real flat field (Figure~\ref{moni}b). Secondly, 200 groups of relative displacements are generated in the same way as the simulated data (1), and the displacements are integrated into the solar observation object (Figure~\ref{moni}a). Thirdly, the displaced observation object is added to a random phase screen with atmospheric seeing $r_0$ equal to 12 cm (Figure~\ref{moni}c), which simulates the distortion of ground-based observed data caused by atmospheric turbulence. Finally, the distorted data are multiplied by the real flat field (Figure~\ref{moni}b) to obtain one group (200 frames) of ground-based observed simulated data, and then the above steps are repeated to generate 100 groups of ground-based observed simulated data.

\subsection{\texorpdfstring{NVST He I 10830\:\AA/H$\alpha$ Observed Data}{NVST He I 10830 Angstrom / H-alpha Observed Data}}

    \label{NVST-data} 

The NVST is a new vacuum solar telescope with an effective aperture of 985 mm \citep{liu2014new}, which is equipped with two groups of terminal instruments: the Multi-band High Resolution Imaging System (MHRIS), and two vertical grating spectrometer systems: Multiband Spectrometers (MBS) for visible spectral lines and High Dispersion spectrometers (HDS) for near-infrared spectral lines \citep{cai2022co}. At present, MHRIS mainly uses three channels: the broadband interference filter channel (TiO-band), and two narrowband Lyot filter channels (H$\alpha$, He I 10830\:\AA). The He I 10830\:\AA\ observation data were recorded with an InGaAs CMOS camera, which was developed by the Astronomical Technology Laboratory of Yunnan Observatory in collaboration with the Kunming Institute of Physics in 2022 \citep{meng2024new}. Some parameters of MHRIS are shown in Table~\ref{T-appendix}.

\begin{table}
\caption{ Some parameters of MHRIS. }
\label{T-appendix}
\begin{tabular}{ccclc}    
                          
  \hline                
Channels & Filters & Bandpass(Å) & Focal length(m) & FOV \\
  \hline
TiO & Interference Filter & 10  & 35  & $180^{\prime\prime}$×$180^{\prime\prime}$ \\
H$\alpha$ & Lyot Filter & 0.5  & 22.5  & $168^{\prime\prime}$×$168^{\prime\prime}$\\
10830\:\AA & Lyot Filter & 0.25  & 20.58 & $96^{\prime\prime}$×$76^{\prime\prime}$\\
  \hline
\end{tabular}
\end{table}

In this paper, the observation data of NVST/He I 10830\:\AA\ and NVST /H$\alpha$ 6563\,\AA\ channels will be used to test the proposed method. The observation times of the two groups of data are 13:42:37 UT on June 5, 2023, and 02:53:42 UT on September 6, 2022, respectively. Among them, the FOV covered by the He I 10830\:\AA\ channel observation data is $96^{\prime\prime}$×$76^{\prime\prime}$ and recorded using a 640× 512 InGaAs CMOS camera, so the scale of each pixel is $0.15^{\prime\prime}$. The FOV covered by the H$\alpha$ (6563\,\AA) channel observation data is $168^{\prime\prime}$×$168^{\prime\prime}$ and recorded using a 2K × 2K CCD camera, so the ratio per pixel is $0.082^{\prime\prime}$. To improve the signal-to-noise ratio (SNR), the data is combined into 1k × 1k, so the scale of each pixel becomes $0.164^{\prime\prime}$ \citep{xiang2016high}.

\section{Method } 
      \label{S-general}      
The method proposed in this paper mainly includes four parts: (1) calculating the correction amount for the average flat field from real-time observation data using the KLL algorithm; (2) setting the initial value; (3) suppressing residual solar structures in the correction amount by group averaging; (4) correcting the average flat field by the calculated correction amount to obtain the real flat field  for the current observation conditions.
\subsection{Calculate the Correction Amount for the Average Flat Field} 
In small FOV ground-based solar observations, the telescope is slightly wobbling due to the wind, causing the solar observation object $s$ to have a displacement $a_i$ between two exposures.

\begin{equation}
s_i(x) = s\left(x - a_i\right)
\end{equation}
where $s_i$ denotes the input signal of the detector, $x$ denotes a pixel on the detector, and $i$ denotes the $i$th frame in a series, \( i =1,2,3...N\). After removing the dark current and amplifier bias at each pixel, we have
\begin{equation}
obs_{i}(x)=s_{i}(x) * f(x)
\end{equation}
where $obs_i$ denotes the the real-time observation data, and \(f(x)\) is the gain of each pixel on the detector (commonly called the flat field)

In order to obtain the real flat field \(f(x)\) of real-time solar observation data more accurately, we split \(f(x)\) into the product of the average flat field and the correction amount, as shown in Equation (3).
\begin{equation}
f(x)=g_{0}(x) * \Delta g(x)
\end{equation} 
where \( g_0(x) \)  is the average flat field, and \( \Delta g(x) \) is the correction amount. We regard \( g_0(x) \) as an initial value, which can be obtained by the multi-frame averaging method \citep{yang2003high}.

Substituting  \( f\left(x\right) = g_0\left(x\right) \ast \Delta g (x) \)  into the Equation (2), the $obs_i$ can be rewritten as
\begin{equation}
obs_{i}(x)=s_{i}(x) * g_{0}(x) * \Delta g(x)
\end{equation}

We follow the optimization strategy improved by \citet{chae2004flat}, where the solar observation object \(s(x)\) is also used as the optimization target. The \(s(x)\) and \( \Delta g(x) \) can be solved using the least square method, as shown in Equation (5). We use the ADAM optimization algorithm \citep{Adam} to jointly solve for \(s(x)\) and \( \Delta g(x) \) in Equation (5). 

\begin{equation}
\begin{aligned}
\left(s(x)^{*}, \Delta g(x)^{*}\right) = 
\mathop{\operatorname{argmin}}_{(s(x), \Delta g(x))} 
\sum_{i=1}^{N}\left\|s_{i}(x) * g_{0}(x) * \Delta g(x) - obs_{i}(x)\right\|_{2}^{2}
\end{aligned}
\end{equation}


\subsection{Set Initial Value}
To accurately solve for the solar observation object \(s(x)\) and the correction amount \( \Delta g(x) \), we set initial values for \(s(x)\) and \( \Delta g(x) \). For setting the initial value of \(s(x)\), it is very important to calculate the relative displacement \( (x_i, y_i) \) of each real-time observation data. The calculation steps are as follows: Firstly, the average frame of the sequence real-time observation data is used as the reference; Secondly, the cross-correlation method is used to calculate the relative displacement \( (x_i, y_i) \) of each real-time observation data with respect to the reference, and the displacements need to be sub-pixel accurate; Finally, each real-time observation data is aligned to the reference, and the average frame after aligning the sequence real-time observation data is used as the initial value \( {s(x)}^0 \) of \(s(x)\). 

Although the above method can obtain \( {s(x)}^0 \) with a high signal-to-noise ratio, which helps to calculate \( \Delta g(x) \) more accurately. However, there are a large number of high-frequency flat field signals in the real-time observation data, such as interference fringes, which will affect the accuracy of relative displacement \( (x_i, y_i) \) calculation. Therefore, we first preprocess the sequence of real-time observation data using the average flat field $g_0\left(x\right)$, and then calculate the relative displacement \( (x_i, y_i) \). Although $g_0\left(x\right)$ can only process part of the flat field signals, the preprocessed sequence real-time observation data is enough to calculate \( {s(x)}^0 \). The initial value of \( \Delta g(x) \) is set to a matrix of ones.

The initial values of \(s(x)\) and \( \Delta g(x) \) are set as shown in Equation (6) - (7).
\begin{equation}
{s\left(x\right)}^0 = \frac{1}{N} \sum_{i}^{N} {{\operatorname{obs}_i\left(x\right)}^\prime / g_0\left(x\right)}
\end{equation}

\begin{equation}
\Delta g(x)^{0}=1
\end{equation}

\subsection{Suppress Residual Solar Structures in the Correction Amount}
Although Equation (5) can calculate the correction amount for the average flat field from real-time observation data, significant solar structures remain in the correction amount. This is because the KLL method assumes that the solar intensity remains stable for a short time, but the small FOV ground-based observed data are severely distorted by atmospheric turbulence. This means that assumptions of the KLL method are violated.

Due to the random distortion of the ground-based observed data caused by atmospheric turbulence, the residual solar structures in the correction amount also exhibit random characteristics. Therefore, we can divide a large number of real-time observation data into multiple groups and then use the Equation (5) to calculate the correction amount from each group of real-time observation data. Finally, multiple groups of correction amount are averaged to suppress the residual solar structures, thus improving the accuracy of the correction amount. 

\begin{figure*} 
    \centering 
    \includegraphics[width=0.8\textwidth]{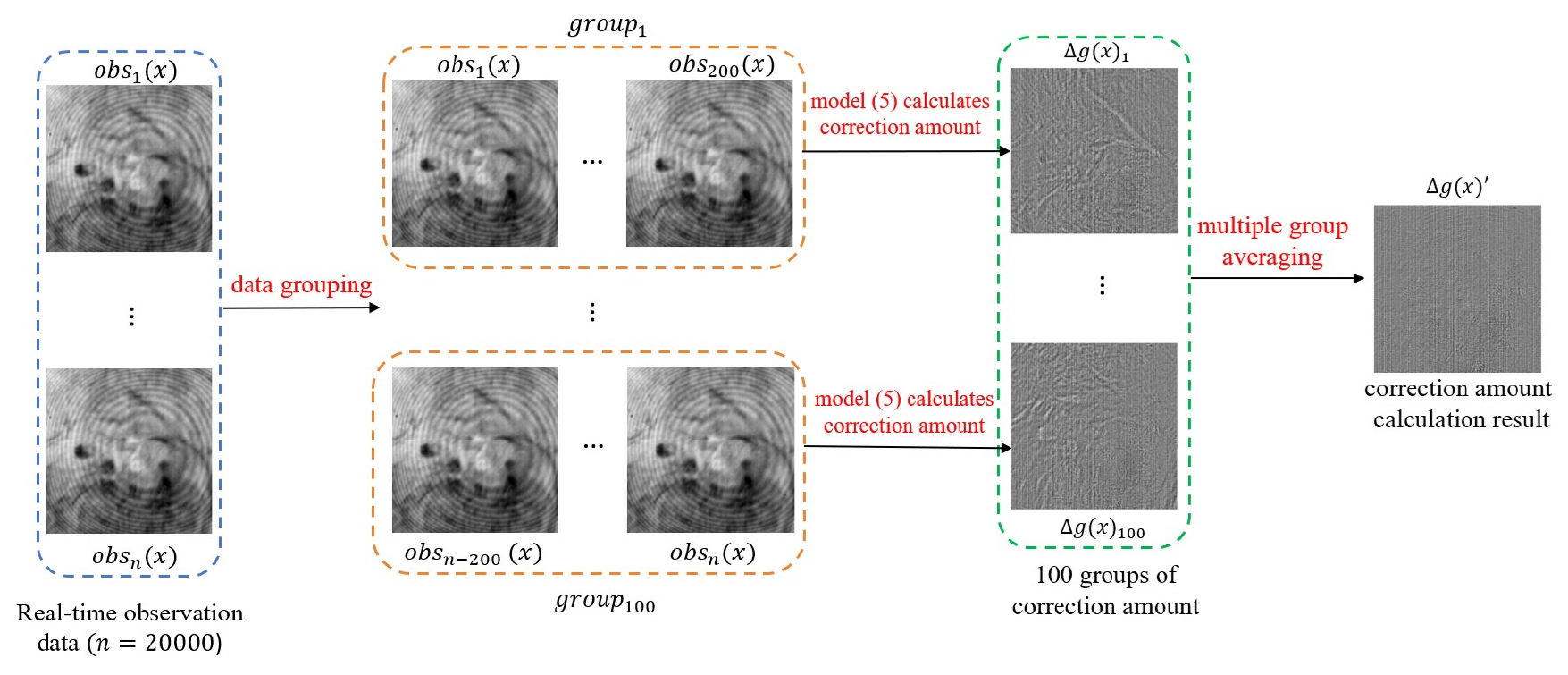} 
    \caption{Flow chart of the algorithm adopts a group averaging strategy to suppress residual solar structures in the correction amount.
}
\label{flow chart}
\end{figure*}

In practical applications, a large amount of real-time observation data is divided into 100 groups (200 frames each) to calculate the correction amount for the average flat field. The algorithm flow chart is shown in Figure~\ref{flow chart}.

We take 200 frames of real-time observation data as a group not only to calculate the correction amount but also to facilitate the subsequent high-resolution reconstruction process. There is no strict limit on the number of real-time observation data frames in each group, but the intensity between each frame of real-time observation data cannot differ too obviously. Otherwise, the correction amount for the average flat field cannot be calculated accurately. 

Considering data processing efficiency, we chose to divide the large amount of real-time observation data into 100 groups. If more groups are added, signal-to-noise ratio of correction amount can be further improved.  

\subsection{Obtain the Real Flat Field for Current Observation Conditions} 
\begin{figure*} 
    \centering 
    \includegraphics[width=0.3\textwidth]{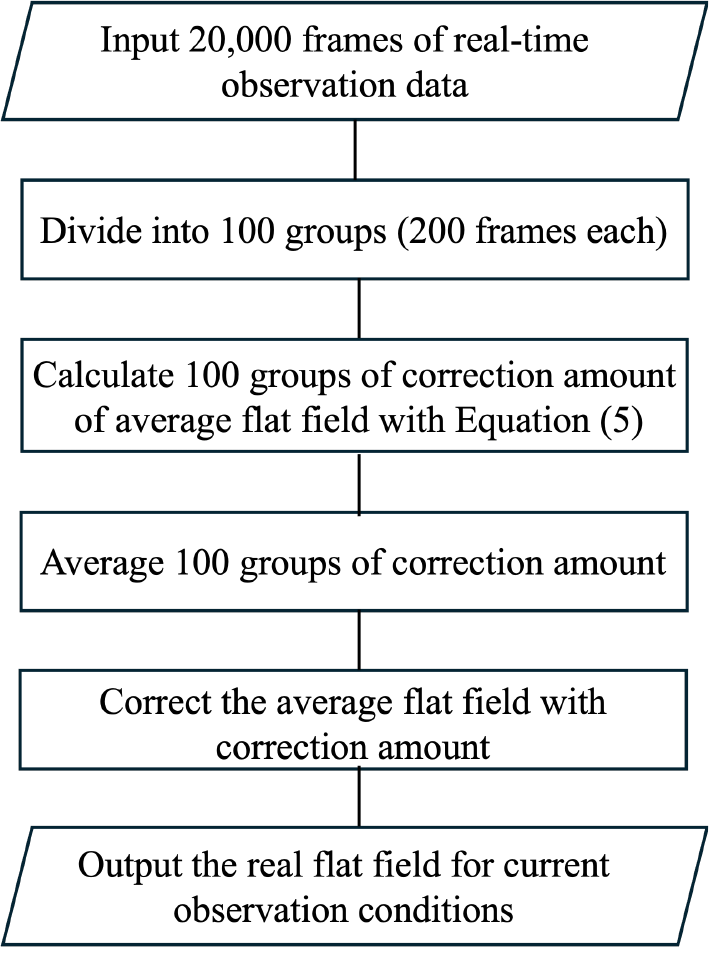} 
    \caption{Flowchart of the overall algorithm.}
\label{Overall}
\end{figure*}

The correction amount $\Delta g'(x)$ for the average flat field calculated in Section 3.3 is used to correct the average flat field $g_0\left(x\right)$, so as to obtain the real flat field $f'(x)$ for the current observation conditions, as shown in Equation (8). The overall algorithm flowchart is shown in Figure~\ref{Overall}.

\begin{equation}
f'(x) = g_0(x) \cdot \Delta g'(x)
\end{equation}
  
\section{Test with Simulated Data} 
      \label{S-features} 
\begin{figure*} 
    \centering 
    \includegraphics[width=0.7\textwidth]{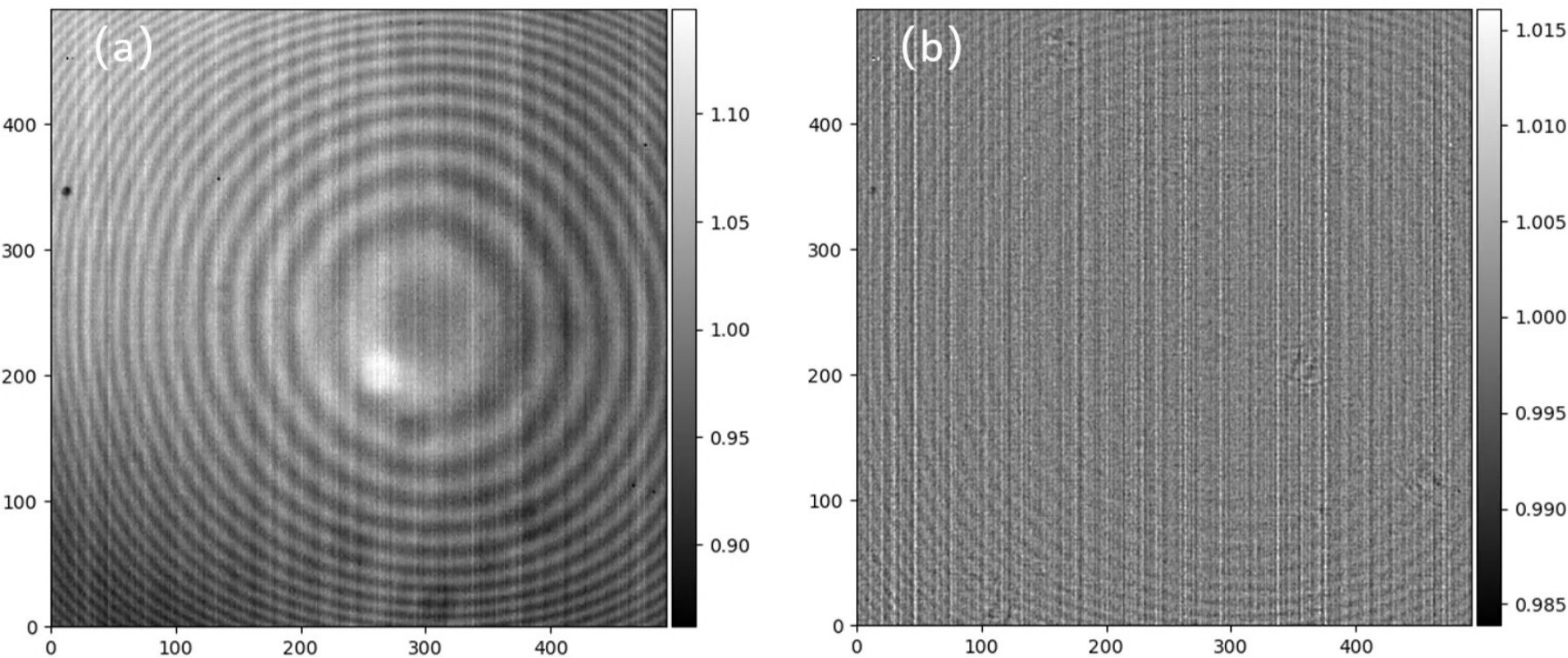} 
    \caption{Panel (a) is the average flat field used for the calculation of the correction amount (observed time is 12:13:42 on June 4, 2023), and panel (b) is the true correction amount, which can be obtained by dividing the real flat field of the simulated data with the average flat field (panel (a)).
}
\label{init_value}
\end{figure*}

In this part, we use the simulated data constructed in Section.~\ref{S-sim} to test the validity of our method in calculating the correction amount for the average flat field. The average flat field (observed time is 12:13:42 seconds on June 4, 2023) is shown in Figure~\ref{init_value}a. Figure~\ref{init_value}b shows the difference between the average flat field (Figure~\ref{init_value}a) and the real flat field (Figure~\ref{moni}b) of the simulated data, which represents the true correction amount for the average flat field. Since we are using simulated data, the true correction amount is known. Therefore, we can evaluate the correction amount calculated by our method. The pixel-to-pixel relative error, $w$ (its proportion reflects the accuracy), and the root mean square error (RMSE) are used as evaluation metrics, and the Equations are shown in (9)-(10). 

\begin{figure*}
    \centering 
    \includegraphics[width=0.9\textwidth]{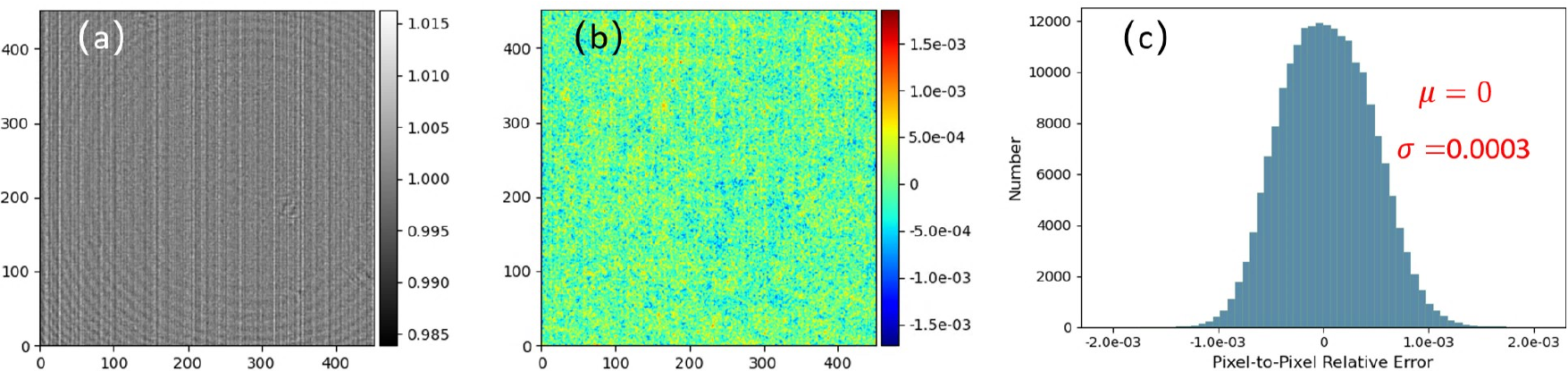} 
    \caption{Visualization of correction amount calculated from space observed simulated data and its error analysis. Panel (a) is the correction amount calculated from the space observed simulated data, panel (b) is the pixel-to-pixel relative error of panel (a), and panel (c) is the error distribution histogram of panel (b).
}
\label{sim-ex-1}
\end{figure*}
\begin{equation}
w=\frac{F(x)-F_{\text {true }}(x)}{F_{\text {true }}(x)} \times 100 \%
\end{equation}
\begin{equation}
RMSE\ =\ \sqrt{\frac{1}{n}\sum_{i=1}^{n}{(F(x_i)-F_{true}(x_i))}^2}     
\end{equation} 
where \(F(x)\) is the calculated correction amount, \( F_{\text{true}}(x) \) is the actual correction amount, and $n$ is the number of pixels. 

Firstly, we use space observed simulated data unaffected by atmospheric turbulence (200 frames) to calculate the correction amount for the average flat field. Figure~\ref{sim-ex-1}a shows the results of the correction amount calculation, and there are no residual solar structures. Figure~\ref{sim-ex-1}b shows the pixel-to-pixel relative error of Figure~\ref{sim-ex-1}a. As can be seen from Figure~\ref{sim-ex-1}b, the correction amount calculated by our method has no obvious errors from fringes or residual solar structures. Specifically, the proportion of $\mid w \mid < 0.1\%$ is about 98.52\%, and the maximum of it is 0.18\%. To visualize the error distribution of Figure~\ref{sim-ex-1}b more intuitively, Figure~\ref{sim-ex-1}c shows the histogram of the error distribution of Figure~\ref{sim-ex-1}b. The mean value of the histogram is 0, and the standard deviation is 0.0003. This indicates that there is almost no systematic error in the calculation result of the correction amount, and the error only fluctuates within a very small range. This demonstrates that our method can efficiently calculate the correction amount for the average flat field from space observed simulated data.


\begin{figure*}
    \centering
    \includegraphics[width=0.9\textwidth]{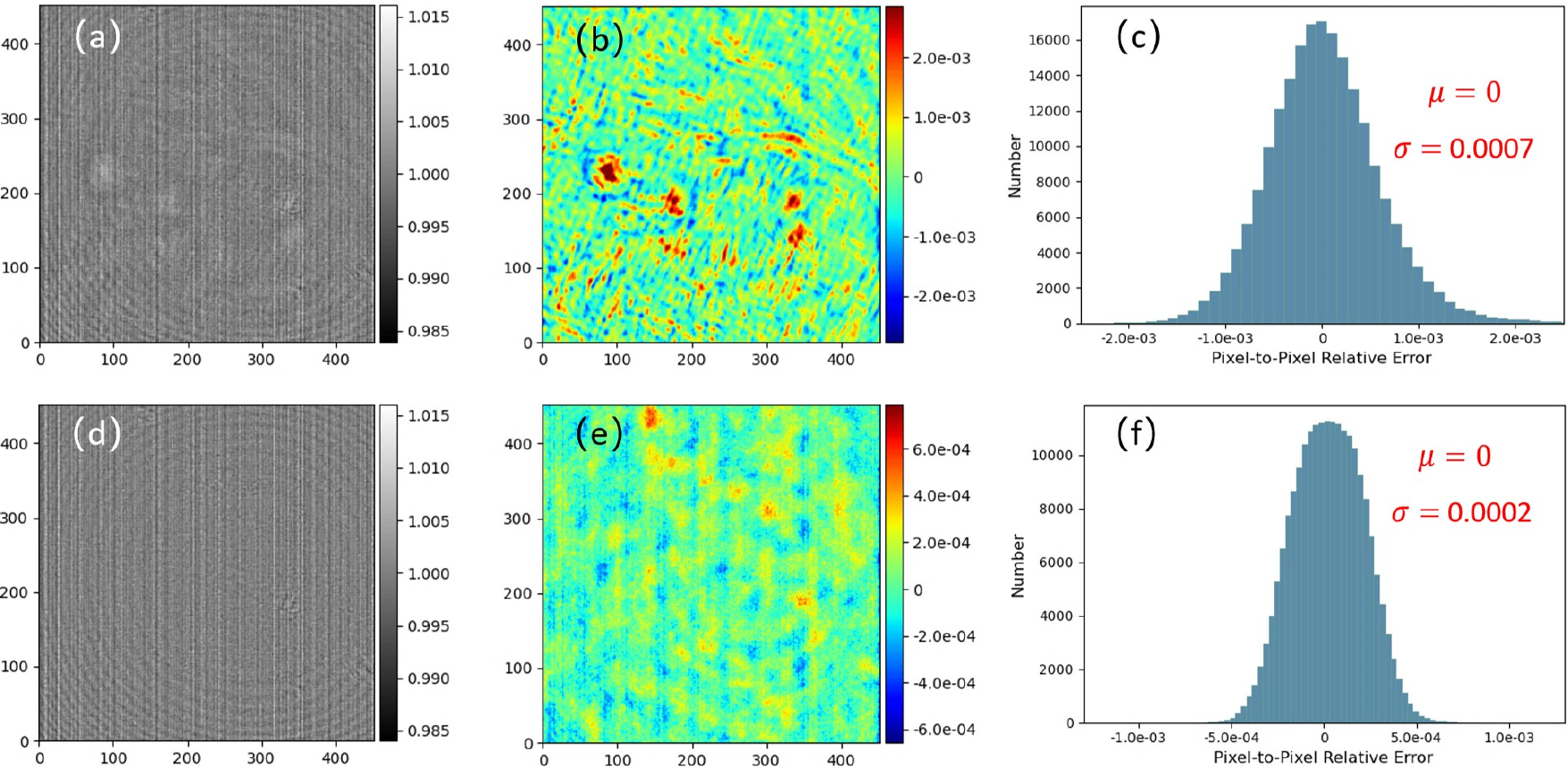}
    \caption{Visual comparison of correction amount calculation results from one group and 100 groups of ground-based observed simulated data and its error analysis. Panel (a)-(c) is the results for one group of ground-based simulated data, and panel (d)-(f) is the results for 100 groups of ground-based simulated data.
}
\label{sim-ex-2}
\end{figure*}

Secondly, we use ground-based observed simulated data affected by atmospheric turbulence to calculate the correction amount for the average flat field. Figure~\ref{sim-ex-2} shows the correction amount calculated from one group and 100 groups of ground-based observed simulated data, and its pixel-to-pixel relative error. Figure~\ref{sim-ex-2}a shows the correction amount calculated from one group (200 frames) of ground-based observed simulated data, and there are obvious residual solar structures. This is because the KLL method assumes that the solar intensity remains stable for a short time, but ground-based observed data is severely distorted by atmospheric turbulence. Figure~\ref{sim-ex-2}b shows the pixel-to-pixel relative error of Figure~\ref{sim-ex-2}a. It can be seen from Figure~\ref{sim-ex-2}b, there is a relatively large error in the residual part of the solar structures. Specifically, the proportion of $\mid w \mid < 0.1\%$ is reduced to 85.70\%, and the maximum error is up to 0.45\%. Figure~\ref{sim-ex-2}c shows the histogram of the error distribution of Figure~\ref{sim-ex-2}b. The standard deviation is up to 0.0007, which is significantly higher than the space observed simulated data. This demonstrates that the residual solar structure caused by atmospheric turbulence reduces the accuracy of the correction amount. 

\begin{figure}
    \centering
    \includegraphics[width=0.45\textwidth]{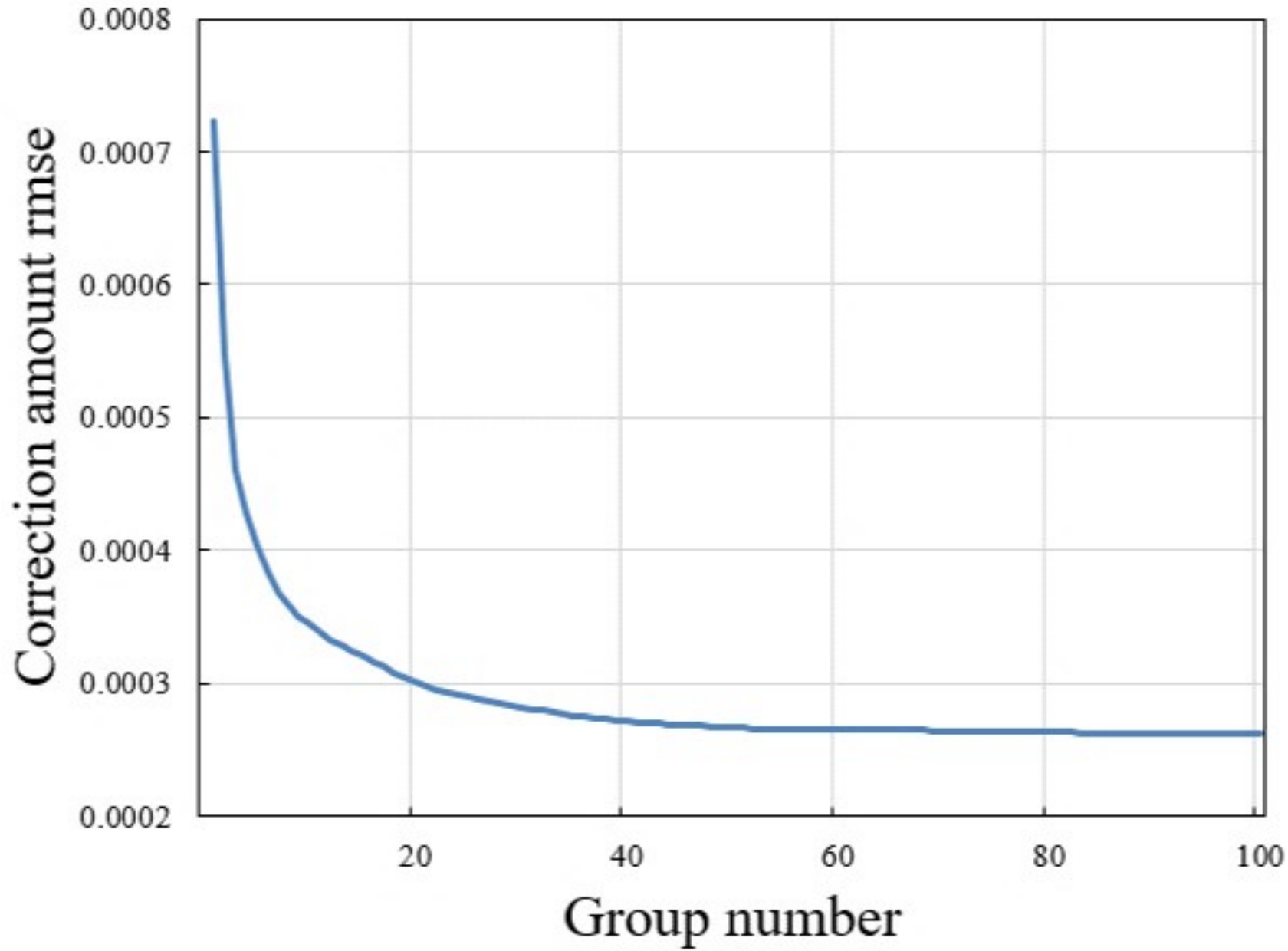} 
    \caption{Root mean square error (RMSE) curve of correction amount calculated from different groups of ground-based observed simulated data.}
    \label{RMSE}
\end{figure}
In order to overcome the effects of atmospheric turbulence, we calculate 100 groups of correction amounts from 100 groups of ground-based observed simulated data. The 100 groups of correction amounts are then averaged to suppress the residual solar structures in the correction amount, thus improving the accuracy of the correction amount. Figure~\ref{sim-ex-2}d shows the average results of 100 groups of correction amount, and it can be intuitively seen that the solar structures are suppressed. Figure~\ref{sim-ex-2}e shows the pixel-to-pixel relative error of Figure~\ref{sim-ex-2}d. Specifically, the proportion of $\mid w \mid < 0.1\%$ increases to 100\%, and the maximum value reduces to 0.082\%. Figure~\ref{sim-ex-2}f shows the histogram of the error distribution of Figure~\ref{sim-ex-2}e. The standard deviation is only 0.0002, which is significantly lower than one group of ground-based simulated data. This demonstrates that grouping the ground-based observed simulated data to calculate the correction amount. Then, the results of multiple groups of correction amount are averaged, which can effectively suppress the residual solar structure in the correction amount, thereby improving the accuracy of the correction amount.

Figure~\ref{RMSE} shows the RMSE curve for the correction amount calculated from different groups of ground-based observed simulated data. The RMSE reflects the overall error of the correction amount. As the number of groups of simulated data increases, the RMSE gradually decreases. When the number of groups reaches 100, the RMSE converges to 0.00026. This further demonstrates the effectiveness of grouped calculation and statistical averaging for determining the correction amount.

\section{\texorpdfstring{Test with NVST He I 10830 Å/H$\alpha$ Observed Data}{Test with NVST He I 10830 Angstrom/H-alpha Observed Data}}
We use NVST He I 10830 Å/H$\alpha$ observed data to test our method. The related parameters are provided in Section.~\ref{NVST-data}. The NVST He I 10830 Å/H$\alpha$ observed data observed time is 13:42:37 UT on June 5, 2023. We first use the average flat field obtained on the same day (observed time is 07:35:26 UT on June 5, 2023) as the initial value. Then, the correction amount for the average flat field is calculated from the NVST/He I 10830 Å observed data. Finally, the calculated correction amount is used to correct the average flat field to obtain the real flat field field for the current observation conditions.
 
\begin{figure*}
    \centering 
    \includegraphics[width=0.72\textwidth]{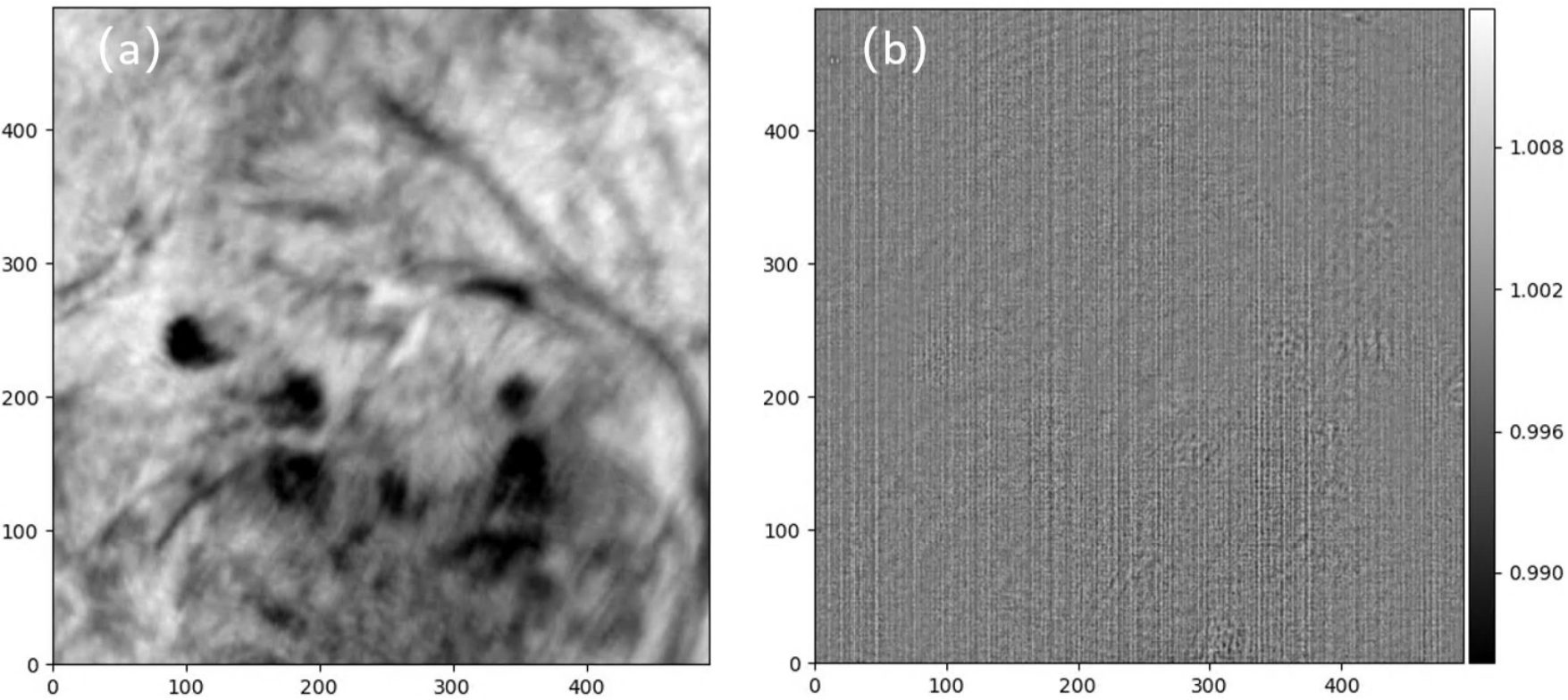} 
    \caption{Panel (a) is the observation object calculated by our method from the NVST/He I 10830 Å  data, and panel (b) is the correction amount for the average flat field calculated by our method.
}\label{correction amount_10830}
\end{figure*}
\begin{figure}
    \centering
      \hspace{-6mm} 
    \includegraphics[width=0.69\textwidth]{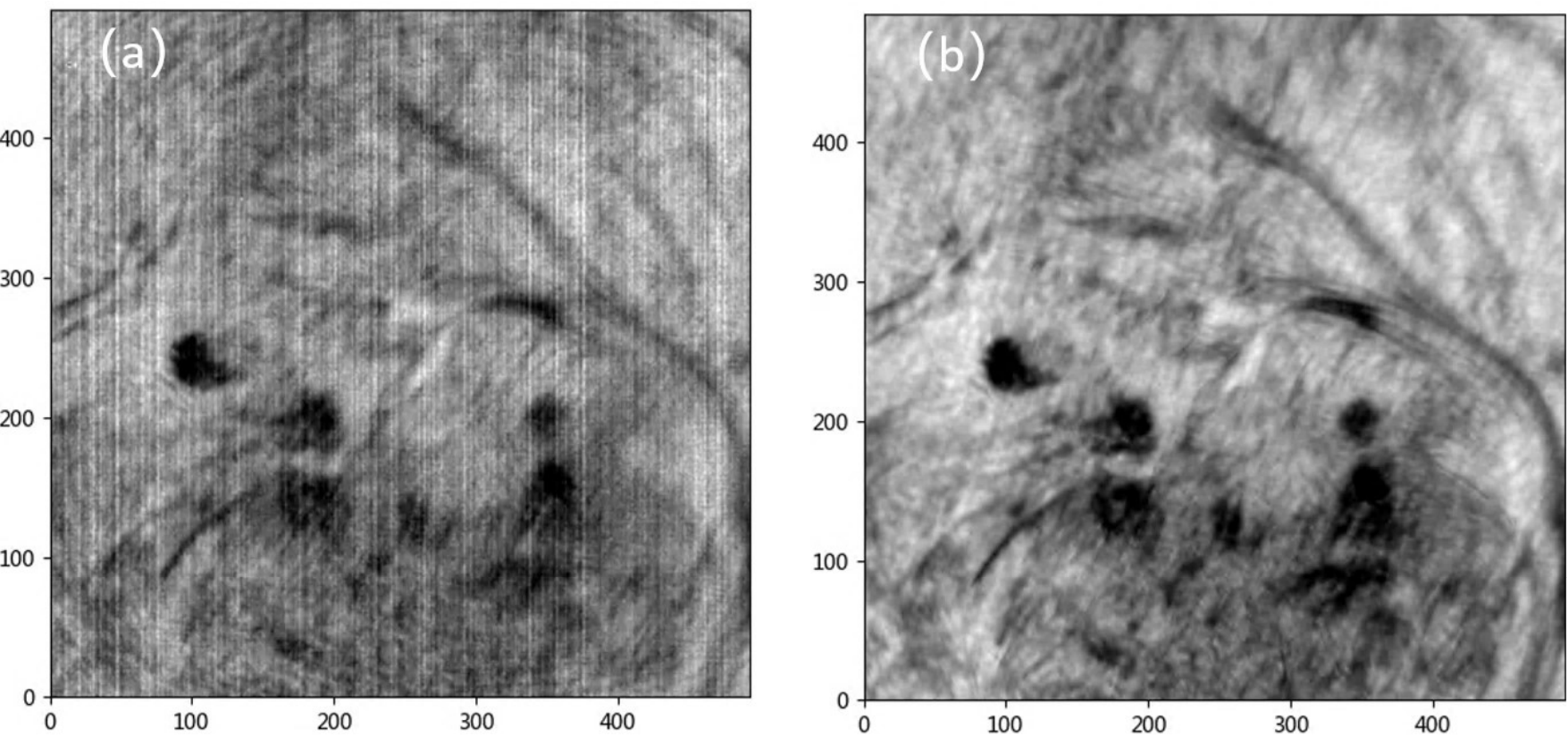} 
    \caption{Comparison of high-resolution data after processing NVST/He I 10830 Å  data with different flat field correct methods. Panel (a) is the high-resolution data after average flat field processing, and panel (b) is the high-resolution data after processing by our method.}
    \label{10830_rec}
\end{figure}
Figure~\ref{correction amount_10830}b shows the correction amount for the average flat field calculated by our method. It can be seen that there are obvious high-frequency signals in the correction amount, such as interference fringes and column-fixed pattern noise. The correction amount reflects the differences between the average flat field and the real flat field of the NVST/He I 10830 Å observed data. It is worth noting that these differences are unavoidable even if the average flat field accuracy is very high. The main reason is that the observed data used to calculate the average flat field are not obtained under actual observation conditions. Instead, we use the real-time observation data as input for the KLL method. The KLL method is then used to dynamically calculate the correction amount for the average flat field. Even if the observation conditions constantly change, we can correct the average flat field by the calculated correction amount so that the obtained flat field can adapt to the current observation conditions. Figure ~\ref{correction amount_10830}a shows the solar observed object calculated by our method, and it can be seen that the observed object has a high signal-to-noise ratio. 

In order to verify the applicability of our method to the NVST/He I 10830 Å observed data, we processed the NVST/He I 10830 Å observed data with our method and the average flat field, respectively, and then reconstructed the processed results using a high-resolution reconstruction algorithm. The reconstruction result is compared as shown in Figure~\ref{10830_rec}. As can be seen from Figure~\ref{10830_rec}a, when the observed data are processed by the average flat field, there are obvious residual signals in the high-resolution reconstruction results, such as interference fringes and column-fixed pattern noise. As shown in Figure~\ref{10830_rec}b, when the observed data are processed using our method, it is difficult to see obvious residual signals in the high-resolution reconstruction result. This demonstrates that the correction amount calculated by our method can effectively correct the average flat field to obtain the real flat field for the current observation conditions, and provide accurate observation data for high-resolution reconstruction.  

\begin{figure}
    \centering
    \hspace{5mm} 
    \includegraphics[width=0.76\textwidth]{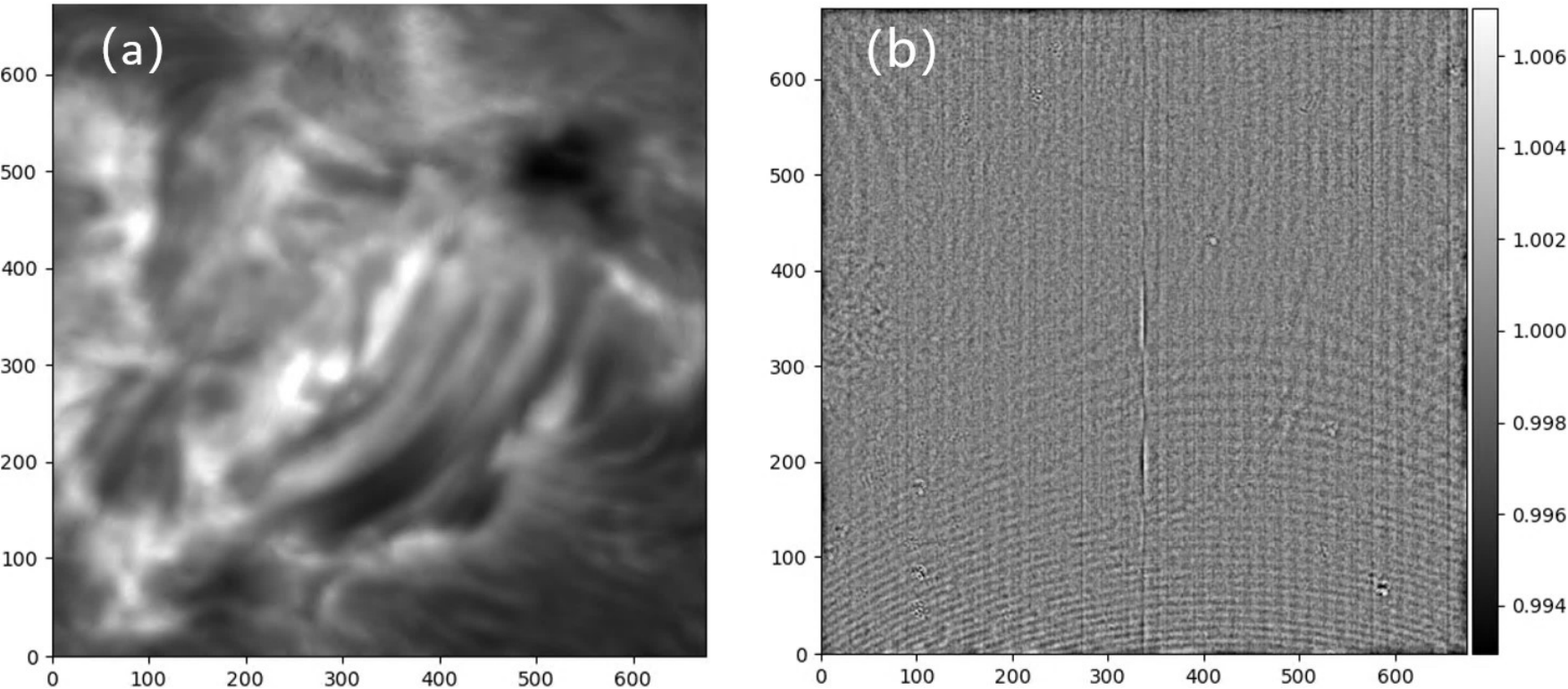} 
    \caption{ Panel (a) is the observed object calculated by our method from the NVST/H$\alpha$ data, and panel (b) is the correction amount for the average flat field calculated by our method.}
    \label{correction amount_ha}
\end{figure}

We also use NVST/H$\alpha$ observation data to test the generalization capability of our method. Figure~\ref{correction amount_ha}b shows the correction amount for the average flat field calculated from NVST/H$\alpha$ observation data, and there are obvious interference fringes in the correction amount calculation result. Since the NVST/H$\alpha$ observation data are obtained using a CCD camera, there is no significant column-fixed pattern noise in the correction amount. Figure~\ref{correction amount_ha}a shows the observed object calculated by our method.
\begin{figure}
    \centering
    \includegraphics[width=0.71\textwidth]{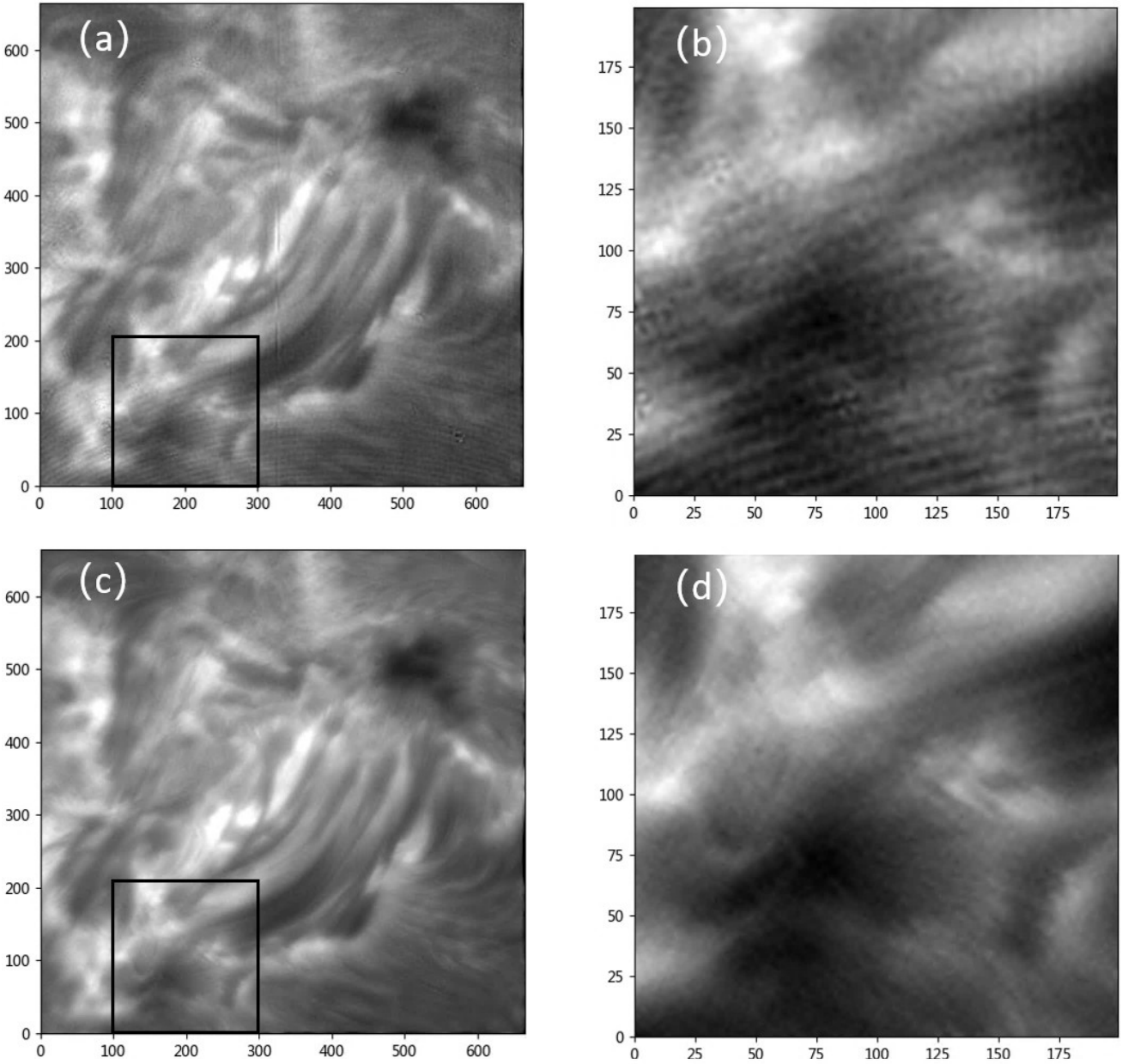} 
    \caption{Comparison of high-resolution data after processing NVST/H$\alpha$  data with different flat field correct methods. Panel (a) is the high-resolution data after average flat field processing, and panel (b) is the high-resolution data after processing by our method, while panel (b) and (d) are the local regions of panel (a) and (c), respectively.
}
\label{ha_rec}
\end{figure}
Like the NVST/He I 10830 Å data, we processed the NVST/H$\alpha$ observed data with our method and the average flat field, respectively, and then reconstructed the processed results using a high-resolution reconstruction algorithm. The reconstruction result is compared as shown in Figure~\ref{ha_rec}. As shown in Figures~\ref{ha_rec}c and d, when the observed data are processed by our method, it is difficult to see obvious residual signals in the high-resolution reconstruction result. In contrast, as shown in Figures~\ref{ha_rec}a and b, when the observed data are processed by the average flat field, there are obvious residual signals in the high-resolution reconstruction result, such as interference fringes. This demonstrates that our method is also applicable to NVST/H$\alpha$ observed data.


\section{Discussion} 

In this part, we mainly discuss the applicability of our method, and the influence of the displacement size of the real-time observation data on the calculation of correction amount for the average flat field. 

The most direct purpose of our method is to obtain the real flat field for the current observation conditions, and improve the quality of subsequent high-resolution reconstruction. In actual observations, the observation conditions constantly change. Therefore, it is necessary that the flat field obtaining method can adapt to the change of observation conditions, and the setting of model parameters cannot rely on the observation conditions. Otherwise, even if reasonable model parameters can be manually adjusted according to the current observation conditions and the real flat field can be calculated, it would be difficult to meet the real-time data processing requirements in actual observations.

Instead, we use an optimization approach to dynamically calculate the correction amount for the average flat field from real-time observation data. The calculated correction amount is then used to correct the average flat field to obtain the real flat field for the current observation conditions. The optimization algorithm can iteratively adjust the calculated correction amount to adapt to changes in the observation conditions. There is no need to manually adjust the model parameters according to different observation conditions. Therefore, our method can effectively correct the observed data under various observation conditions and support the application of all-day observation. In addition, we use GPU to process the grouped real-time observation data, which accelerates the flat field calculation. It only takes about five minutes to obtain the real flat field for the current observation conditions. Therefore, our method has good applicability.

The displacement size of the real-time observation data is very important in calculating the correction amount for the average flat field. In actual observations, the displacement size of real-time observation data is small. However, these small shifted data can effectively capture high-frequency correction amounts, such as interference fringes and column-fixed pattern noise. In fact, these signals are the main factors affecting the quality of high-resolution data. We use the calculated high-frequency correction amount to correct the average flat field, which greatly reduces the residual signal in the observed data after flat field correction, thereby improving the high-resolution reconstruction results. Since the low-frequency flat field signal of the observed data is relatively stable in a short time, the average flat field as the initial value can process most of the low-frequency flat field signals.
      
\section{Conclusion} 
In this paper, we propose a method of extracting the flat field from the real-time solar observation data, which can obtain the real flat field for the current observation conditions. In our method, a correction amount is calculated from real-time observation data using the optimization approach. Then, the calculated correction amount is used to correct the average flat field. Furthermore, a large number of real-time observation data are grouped to calculate correction amounts, and then multiple groups of correction amounts are averaged. This strategy overcomes the residual solar structures in the correction amount calculation results caused by atmospheric turbulence, further improving the accuracy of the correction amount. The test results of space and ground-based simulated data demonstrate that our method can effectively calculate the correction amount for the average flat field. The real NVST 10830\:\AA/H$\alpha$ observation data confirm that the calculated correction amount effectively corrects the average flat field to obtain the real flat field for the current observation conditions. We want to highlight that the proposed method applies not only to chromosphere data but also to photosphere data. 

However, it is difficult to capture the low-frequency correction amount using small displacement data. In the future, we can increase the telescope's wobble to introduce larger displacement, which will better capture the low-frequency correction amount and further improve the accuracy of the correction amount calculation.

\begin{acks}
The author would like to thank the astronomical technology laboratory of Yunnan astronomical observatory research institute for providing observation data and support for this research.\end{acks}

\bibliographystyle{spr-mp-sola}

\end{document}